\documentclass[10pt,conference]{IEEEtran}
\IEEEoverridecommandlockouts
\usepackage{cite}
\usepackage{amsmath,amssymb,amsfonts}
\usepackage{algorithmic}
\usepackage{graphicx}
\usepackage{textcomp}
\usepackage{xcolor}
\def\BibTeX{{\rm B\kern-.05em{\sc i\kern-.025em b}\kern-.08em
    T\kern-.1667em\lower.7ex\hbox{E}\kern-.125emX}}

\newcommand\vip[1]{\texttt{#1}}
\newcommand\service[1]{\textit{#1}}

\begin{document}

\title{Automating chaos experiments in production}

\author{
\IEEEauthorblockN{Ali Basiri}
\IEEEauthorblockA{
\textit{Netflix}\\
Los Gatos, CA \\
abasiri@netflix.com}
\and
\IEEEauthorblockN{Lorin Hochstein}
\IEEEauthorblockA{\textit{Netflix}\\
Los Gatos, CA \\
lhochstein@netflix.com}
\and
\IEEEauthorblockN{Nora Jones}
\IEEEauthorblockA{\textit{Netflix}\\
Los Gatos, CA \\
noraj@netflix.com}
\and
\IEEEauthorblockN{Haley Tucker}
\IEEEauthorblockA{\textit{Netflix}\\
Los Gatos, CA \\
haley@netflix.com}
}

\maketitle

\begin{abstract}
Distributed systems often face transient errors and localized component degradation
and failure. Verifying that the overall system remains healthy in the face of
such failures is challenging. At Netflix, we have built a platform for automatically
generating and executing chaos experiments, which check how well the production system can
handle component failures and slowdowns. This paper describes the platform and our experiences
operating it.
\end{abstract}


\section{Introduction}

Consider a service delivered to users over the Internet. All such services are implemented
as distributed systems. The smallest such service would involve two machines (a single client and server),
while the larger ones, such as Netflix, are composed of thousands of servers.

Large-scale distributed systems contain many failure modes, as there are many
opportunities for individual component failures and unexpected interactions
among components\cite{Gunawi2014}. To ensure that a system remains available to users,
engineers build resiliency into the system through strategies such as timeouts,
retries, and fallbacks \cite{release-it}.

Ideally, a distributed system will degrade gracefully if an individual component runs
into trouble. In general, it is difficult to know whether engineering resiliency mechanisms
will actually work to keep the overall system healthy if some part of the system goes bad.

\textit{Chaos engineering} \cite{chaoseng, chaosengbook} is an emerging approach in industry to evaluate the
resiliency of a distributed system by running experiments against the system, in production. These
experiments can identify weaknesses that could lead to outages if left unchecked.

At Netflix, we have developed a platform to automate the generation and execution of chaos
experiments. These experiments run directly against our production environment. The
authors are all members of the Resilience Engineering team, which is a centralized
team that develops chaos engineering tools. In this paper, we discuss
the design of our system and our experiences operating it. This platform was built in
about three years by a team whose size varied between one and four software engineers.

\section{Context: Netflix}

\subsection{Overview}

Netflix is a service that enables customers to stream television shows and movies over the Internet,
on a variety of different types of devices, such as smart TVs, set-top boxes,
smartphones, tablets, laptops, and game consoles.

One of the most important key performance indicators for Netflix is \textit{availability}. 
In our context, we consider the system to be available if users are successfully
able to stream video on their devices. We usually express service availability
as a percentage of users who are successfully able to stream video, over some
interval of time. For example: 99.99\% availability (``four nines'') means, roughly, that
99.99\% of the time, when a user tried to start watching a video, they were successful.

While Netflix does not have the same availability requirements as, say, a telecommunications
company, availability is still important for the business: if customers experience
service interruptions, they are more likely to cancel their subscriptions.


\subsection{Microservice architecture} 

\begin{figure*}[tbp]
\includegraphics[width=\textwidth]{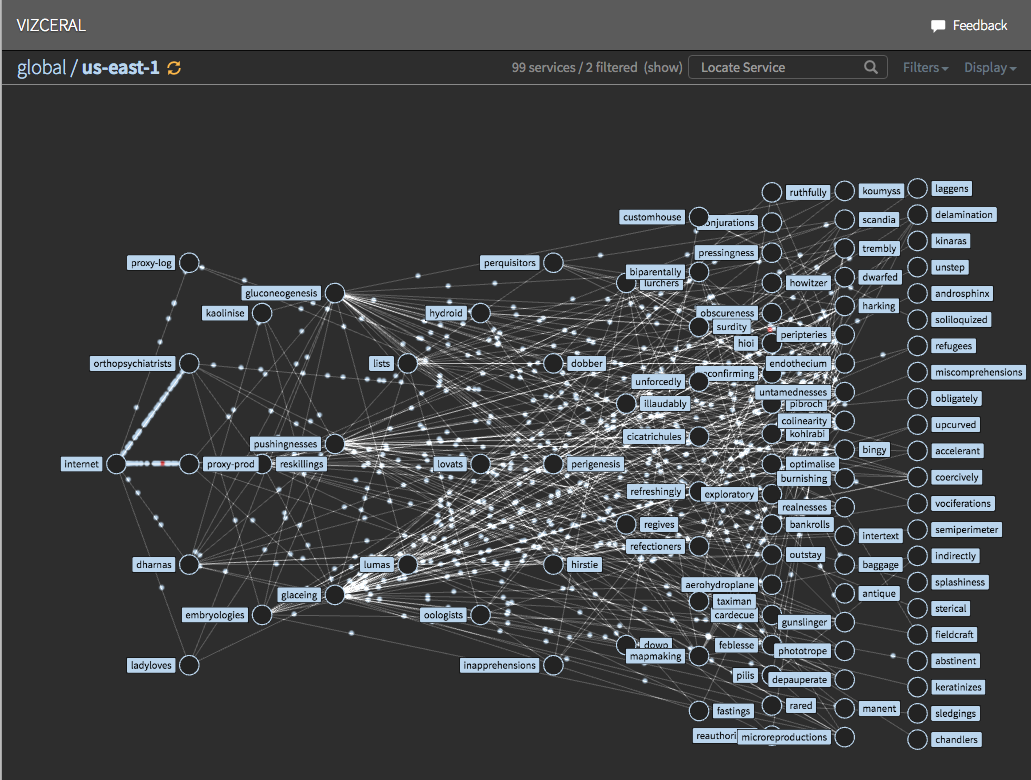}
\caption{A visualization of a microservice architecture}
\label{fig-architecture}
\end{figure*}

From the perspective of the end-user, Netflix is a single service. Internally,
Netflix is implemented using a microservice architecture \cite{microservices}:
a collection of services\footnote{For the remainder of this paper, we use the
term \textit{service} to refer to a microservice.} that communicate with each other via remote procedure
call (RPC). Fig.~\ref{fig-architecture} shows a Vizceral\cite{vizceral} visualization of a microservice architecture,
where each node represents a cluster of servers that make up a single service, and requests, referred to
collectively as \textit{traffic}, flow through the system from left to right.

\begin{figure}[tbp]
\includegraphics[width=\columnwidth]{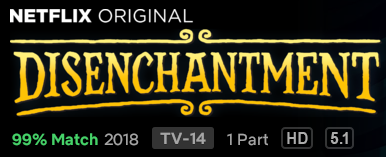}
\caption{Screen shot of the top-left corner Netflix UI. Note that information such as the match score (99\% Match),
 and the badges which show video and audio metadata (HD, 5.1) are retrieved from different microservices.}
\label{fig-ui}
\end{figure}

Many features that Netflix exposes to users are localized to specific services.
For example, Netflix allows users to search for a specific title: this functionality
is implemented by the \textit{search} service. Fig.~\ref{fig-ui}, which shows a screen shot
of part of the Netflix user interface, demonstrates how the user interface is assembled from the
output of different microservices. There is one service that is responsible for presenting the match score,
another one for showing metadata  about the video stream (HD, 5.1), and another one that will process the results if 
a user clicks the "Rate this title" option and give the title a thumbs up or thumbs down.

Netflix uses a microservice architecture to improve velocity: services are owned by different teams,
and each team can deploy a service independently, without needing to coordinate with other teams. A microservice
architecture can also increase availability by reducing the size of fault domains:
if one service gets into a bad state, it doesn't necessarily put the overall system into a bad state. For example,
if the service that processes user rating fails, the overall system should handle that failure gracefully and the user
should not even notice that this service has failed.

\subsection{Resilience through timeouts, retries, and fallbacks}

The Netflix control plane operates in an unforgiving environment: one of constant
change within a public cloud. In this type of environment, there are many potential
sources of failure, stemming from the infrastructure itself (e.g., degraded hardware, transient networking problem) or,
more often, because of some change deployed by Netflix engineers that did not have the intended effect.

Three of the strategies that Netflix employs to achieve resilience are \textit{timeouts},
\textit{retries},  and \textit{fallbacks}. All RPCs are configured with timeouts.
An RPC might time out for a number of reasons, including problems with the particular
server being called (e.g., server is overloaded) or problems with the networking infrastructure.

Many failures are either transient or isolated to a specific server, which means that a retry to
a different server can often resolve the situation. However, not all problems are 
transient: if a bad code push results in a downstream service being in a bad state (e.g.,
returning errors for all requests), then a retry will not resolve the situation. In these
cases, we rely on \textit{fallbacks}, which is a sensible default response. For example, if we cannot suggest a row of viewables
that are personalized for an individual user, we can fall back by serving a row of viewables that
are not personalized (e.g., what is currently popular on Netflix). Fallbacks enable the
system to degrade gracefully when encountering localized failures.

Many Netflix services use the Java-based Hystrix\cite{hystrix} library to implement fallbacks and the circuit breaker pattern\cite{release-it}.
The Hystrix library has the notion of \textit{commands}. A Hystrix command is a wrapper around logic
that can potentially fail. A common Hystrix use case is to wrap a call for an RPC client. 
The Hystrix command can be configured to support timeouts, and fallbacks to handle errors. Hystrix commands are multithreaded:
by default, a Hystrix command is associated with a threadpool with ten threads. 

One of the challenges with timeouts and fallbacks is that these behaviors are not exercised
as frequently as the \textit{happy path}, which means we have less confidence they will work as expected.
This is one of the key motivations for the development of the Chaos Automation Platform.

\section{ChAP}

\subsection{Overview}

We have developed a system called the Chaos Automation Platform (ChAP)\cite{issre} for
running chaos engineering experiments within the Netflix microservice architecture.

Most ChAP experiments focus on evaluating whether the overall system would remain healthy
if one of the services degraded. Two failure modes we focus on are a service becoming
slower (increase in response latency) or a service failing outright (returning errors).
The service-level view of failure is useful because many different types of faults can be modeled
as a service slowing down or returning errors. In particular, many bad code pushes
(deployment of code with defects) can be modeled as a service that returns errors,
and many forms of resource exhaustion (e.g., CPU, threads, memory, network bandwidth) can
be modeled as a service slowing down.

ChAP is effectively an orchestration system that interacts with a number of internal
Netflix services in order to carry out experiments.  ChAP leverages a fault\footnote{
In this paper, we sometimes use \textit{fault} and \textit{failure} interchangeably, which
is consistent with the Chaos Engineering industry community usage.} injection
system developed inside of Netflix called FIT\cite{fit},
which does fault injection at the application level. FIT takes advantage of the fact that
applications deployed inside of the Netflix control plane use a common set of Java libraries.
These libraries have hooks in them that enable us to inject faults at runtime.

Fault injection is typically done by annotating incoming requests with metadata that
indicate that a call should be failed.  This metadata is passed along as requests
propagate through the system.

FIT supports two types of fault injection:

\begin{itemize}
    \item failure - throw an exception instead of executing the call
    \item latency - add latency before executing the call
\end{itemize}

FIT supports injection into a number of different libraries, including RPC clients (both REST
and gRPC), Hystrix\cite{hystrix}, EVcache\cite{evcache}, and Cassandra database clients.

\subsection{Motivating example: failing the bookmarks service}

To understand what ChAP does, we'll walk through an example experiment. There is
a service, which we'll call \service{bookmarks}, which is responsible for keeping track
of where a user when they previously watched a video. For example, if you previously
watched 45 minutes of the romantic comedy ``Set it up'', quit the Netflix app, and then
returned to watch, the bookmarks service is responsible for identifying the
location in the video to continue watching from. 

%
%

The bookmarks service adds value for users, but it is not essential to the functioning of the overall system.
If the bookmarks service failed, then we would still expect that users would be able to use
the service and watch videos: they simply wouldn't be able to continue from the last place
where they watched.

Our hypothesis is that users should still be able to stream video when the bookmarks service fails.
We can test this hypothesis by running an experiment where we intentionally cause the
bookmarks service to fail for a sample of the user population, and verify that this sample of users
are still able to stream videos successfully.

Although ChAP supports both server-side and client-side RPC fault injection, in our experiments we generally
inject faults into the RPC clients: the service that is the subject of the experiment is the caller upstream
of the service that we intend to fail. In this example, the API service\cite{api} is the upstream that 
calls the bookmarks service.

\begin{figure*}[tbp]
\includegraphics[width=\textwidth]{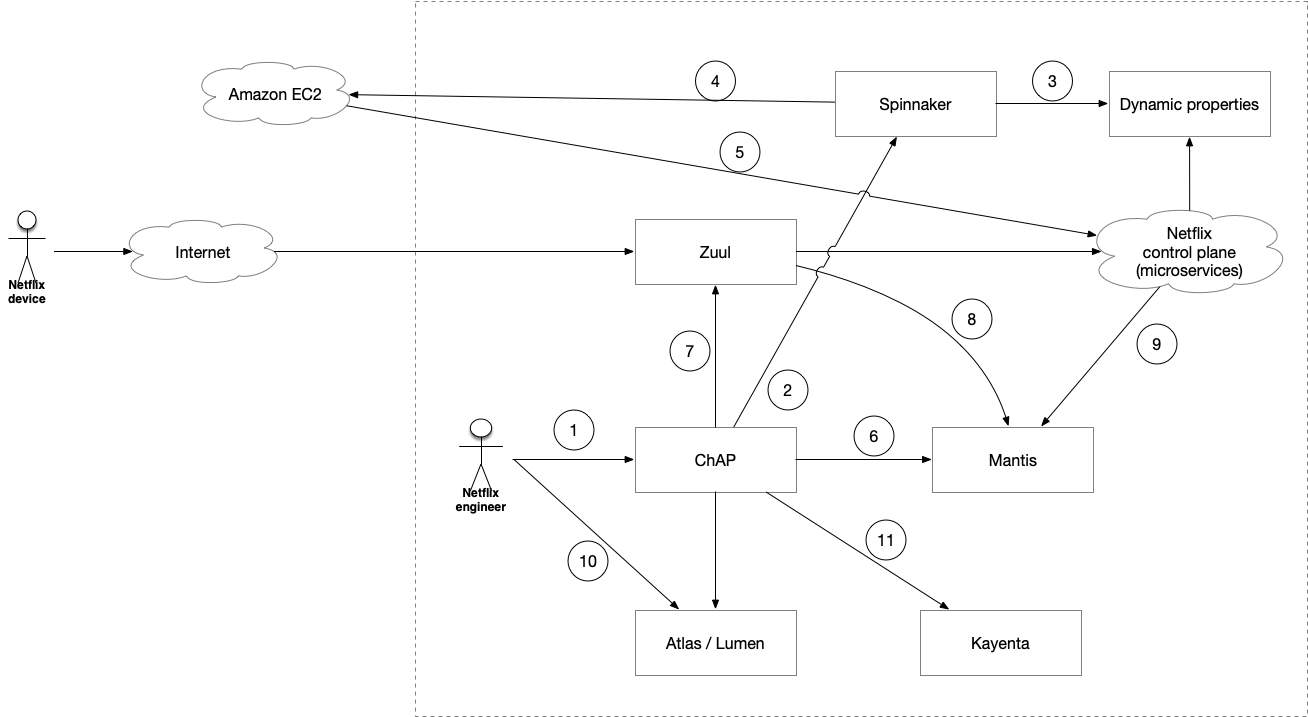}
\caption{ChAP service interactions. The dashed rectangle shows the system boundary of interest: elements outside of the boundary
(e.g., Amazon EC2) are outside of the system. For the purposes of this diagram, we considered Netflix devices as outside
of the system. The callouts (e.g. \textcircled{1}) are explained in the text.}
\label{fig-system-diagram}
\end{figure*}

\begin{figure*}[tbp]
\includegraphics[width=\textwidth]{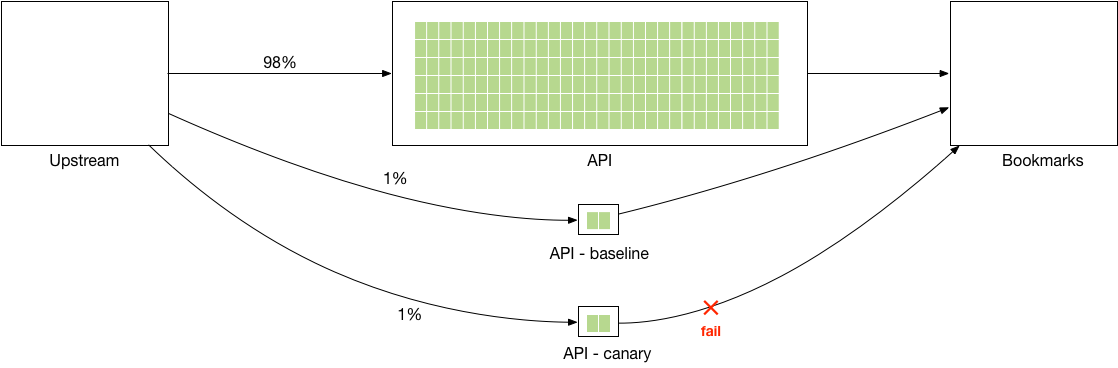}
\caption{ChAP experiment injecting failure into bookmarks from API}
\label{fig-bookmarks}
\end{figure*}

For this example, we'll assume that the API cluster contains 180 servers. We also assume
that the experiment will impact 1\% of users. This means that we will select randomly select
1\% of active users to be in the experiment (treatment) group. We will also randomly select 1\% of users
to be in the control group. Fig.~\ref{fig-system-diagram} depicts some of the internal Netflix services that ChAP
must orchestrate to run an experiment. A ChAP experiment in progress is depicted in Fig.~\ref{fig-bookmarks}.

\subsection{Define the experiment}

The user, a Netflix engineer, creates an experiment using the ChAP UI \textcircled{1}. In this case, the experiment is: \textit{fail RPCs to the bookmark service
and verify the system in general and the API service in particular remain healthy.}

The user specifies:

\begin{itemize}
    \item Failure scenario: fail calls to the bookmark service
    \item Service to observe: API
\end{itemize}

\subsection{Create the baseline and canary clusters}
ChAP first calls out to Spinnaker\cite{spinnaker} \textcircled{2}. Spinnaker is Netflix's continuous
delivery system. Spinnaker serves as an interface for two important services: Amazon EC2 and Netflix's
internal dynamic properties system.

Netflix engineers use Spinnaker's web-based user interface (UI) to define deployment pipelines for deploying
their services onto Amazon EC2.  Netflix engineers also use Spinnaker's UI as an interface into Netflix's internal dynamic properties system.
Netflix services support dynamic configuration: configuration properties can be changed at runtime through Spinnaker.

ChAP uses Spinnaker to make requests against Amazon EC2 \textcircled{4} to provision two smaller copies of the API cluster inside of the Netflix control plane \textcircled{5} . These clusters are referred to as the ``baseline'' and ``canary''
clusters. Traffic from users in the control group will be routed to the baseline cluster, and traffic from users in the experiment group
will be routed to the canary cluster. We borrow the baseline/canary terminology
from canary deployments\cite{continuous-delivery}. Spinnaker is also used to copy any dynamic configuration
settings from the original cluster to the baseline and canary \textcircled{3}. The dynamic configuration settings are copied
before creating the new clusters so that the settings are in place before the new instances boot.

We size the baseline and canary clusters to be 1\% of the size of the original cluster, in this example
that corresponds to two servers per cluster. 

Netflix services use routing identifiers called VIPs (virtual IPs)\cite{release-it}. VIPs are strings which behave similarly to DNS hostnames. Servers
advertise VIPs to Eureka\cite{eureka}, the service discovery service. RPC clients query Eureka
to translate a VIP into a list of IP addresses.  The baseline and canary clusters are each configured to advertise a VIP
that is different from the original cluster. For example, if the servers in the original
API cluster advertise the VIP \vip{api}, then the baseline cluster would be assigned the VIP \vip{api-chap-baseline}
and the canary cluster would be assigned the VIP \vip{api-chap-canary}. Because the RPC clients upstream of API
are only configured to call out to the \vip{api} VIP, the baseline and canary clusters do not receive any traffic
when they first come up.

\subsection{Start low-latency monitoring job}

As will be discussed in more detail in Section~\ref{sec:dashboard}, ChAP relies on the Atlas\cite{atlas} telemetry system as a source
for many of the dashboard graphs as well as the final analysis of the experimental
results.

However, the most recent data we can reliably query from Atlas is typically five minutes old. If
an experiment has revealed a significant vulnerability and customers are
being severely impacted, we wish to detect this as soon as possible so we can abort the experiment.

In order to obtain lower latency telemetry data on critical business metrics that indicate whether customers
are being impacted, we rely on an internal stream processing system called Mantis\cite{mantis}.

Mantis is a platform that allows Netflix engineers to define jobs that consume events that are generated
by different microservices.  ChAP starts a Mantis\cite{mantis} job \textcircled{6} which keeps a count of the number of successful and failed video
start play and download events, for users in both the baseline and canary groups. The Mantis job
sends this data back to ChAP once a second. The latency of this data is on the order of seconds rather
than minutes.

\subsection{Sample from users, configure routing and fault injection}

ChAP publishes an event through an internal data publishing and subscription (pub/sub) service to indicate that the experiment should begin.

The event contains the following information:

\vspace{10pt}

\begin{tabular}{l|p{4cm}}
\multicolumn{1}{c}{Data} & \multicolumn{1}{c}{Example} \\
\hline 
Experiment size & 1\% of users \\
Failure to inject & fail calls to the \textit{bookmarks} service \\
VIP of original cluster & \vip{api} \\
VIP of baseline cluster & \vip{api-baseline} \\
VIP of canary cluster & \vip{api-canary} \\
\end{tabular}
\vspace{10pt}


Zuul\cite{zuul} is the \textit{front door} to the Netflix control plane. It is a reverse-proxy service
that receives all inbound requests from Netflix devices and routes the traffic into the control plane.
Zuul also supports the notion of \textit{filters}, which are functions that can be added to Zuul
in order to perform processing on incoming requests. The Resilience Engineering team owns a Zuul
filter which provides the functionality required for doing ChAP experiments, as described below:

The ChAP Zuul filter receives the event published by ChAP \textcircled{7} and randomly assigns 1\% of
users to the baseline group, and 1\% of users to the canary group. 

If an end-user is selected to be part of the baseline group, then all requests associated with that user will be annotated with a header with the following routing rule:  \vip{api}~$\rightarrow$~\vip{api-baseline}.

Any RPC client that is configured to make calls to the \vip{api} VIP
will instead make the call to \texttt{api-baseline} VIP. This will ensure that the users in
the baseline group have their traffic routed to the api-baseline cluster.

If an end-user is selected to be part of the canary group, then all requests associated with
that user will be annotated with a header with the following routing rule:  \vip{api}~$\rightarrow$~\vip{api-canary}

In addition, all requests will be annotated with a header with the following
fault injection rule: \textit{fail calls to the bookmark service}. Any RPC client configured to
call the bookmarks service will instead return an error. 

Zuul emits a Mantis event for each request. When a request involves an end-user that has been
selected to be a member of the baseline or canary group, this event is consumed by the ChAP Mantis job \textcircled{8},
which keeps track of which end-users are in the baseline and canary groups.

Other services in the Netflix control plane emit events that are associated with ``start play'' events (successes, errors).
These events are also consumed by the ChAP Mantis job: if a relevant event is associated with a user in 
a baseline or canary group, then the relevant metric counter is incremented.

\subsection{Display dashboard}
\label{sec:dashboard}

Once the experiment starts, the user is presented with a dashboard
which plots relevant metrics \textcircled{10}. Netflix relies on two internal system for generating
operational dashboards: Atlas and Lumen.

Atlas\cite{atlas} is a telemetry platform for collecting and graphing time-series data in order
to provide operational insight. Netflix engineers often interact with Atlas
using a user interface called AtlasUI that allows users to generate graphs using
either a graphical user interface or through a custom query language called Atlas
Stack Language.

AtlasUI is useful for ad hoc generation of individual graphs, but engineers typically
want to view a collection of previously specified graphs on a single page. Lumen\cite{lumen}
is a platform that allows Netflix engineers to generate dashboards based on
graphs generated by the Atlas back-end.

ChAP uses Lumen dashboards for displaying metrics data.
This data is used to determine whether or not the injected faults had a negative
impact on system health.

There are two main classes of metrics we use to determine whether the injected fault
has had a negative impact on system health:

\begin{itemize}
    \item key performance indicators (KPIs) for the users in the canary group versus the baseline group
    \item health metrics for the canary cluster versus the baseline cluster
\end{itemize}

\begin{figure}[tbp]
\includegraphics[width=\columnwidth]{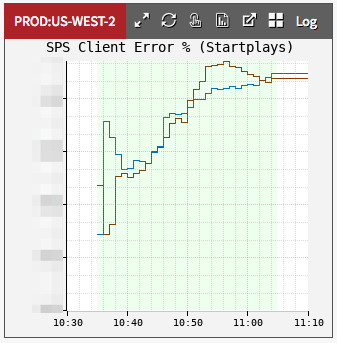}
\caption{Example graph from the dashboard generated by ChAP. This one shows one of the KPIs for the users in the baseline (control)
and experiment (canary) group. The specific KPI shown here is the cumulative percentage of startplay errors reported by the client device.
The area shaded in green indicates the time when fault injection is active.  The y axis is intentionally obscured here to hide proprietary information.}
\label{fig-kpis}
\end{figure}

The key performance indicators we are concerned with are around the ability of users to be
able to stream video. In particular, we track a count of stream-starts per second (SPS), which is
a count of the number of successful video stream starts. We collect SPS counts from both the
server side and the device side, and we also track SPS errors, as shown in Fig.~\ref{fig-kpis}. Note
that because error counts are generally much lower than success counts, it is much more likely that
the error counts differ between the two groups due to random variation.

\begin{figure}[tbp]
\includegraphics[width=\columnwidth]{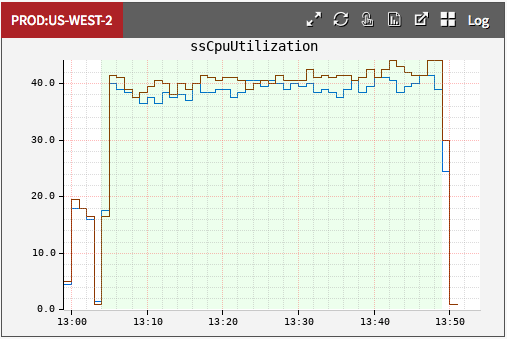}
\caption{CPU utilization (\%) is one of health metrics tracked in experiments.}
\label{fig-server-stats}
\end{figure}

We also compare health metrics between the baseline and canary deployments. We examine metrics
such as request rate, latency, error rate, and CPU utilization. One example, CPU utilization,
is shown in  in Fig.~\ref{fig-server-stats}.

\subsection{Monitor ongoing experiment}

ChAP monitors the experiment as it runs to verify that there
is no significant negative impact to customers. See Section~\ref{sec-safety} for more details.

\subsection{Cleanup}

Once the experiment has completed, ChAP unpublishes the experiment event
to stop Zuul from re-routing traffic and injecting faults. ChAP makes calls against Spinnaker to tear
down the baseline and canary clusters.

\subsection{Analyze the results}

Finally, ChAP calls out to Kayenta\cite{kayenta} \textcircled{11}, Netflix's automated canary
analysis system. Netflix engineers often use canary deployments to verify that new code
being pushed to production has not introduced a regression. Kayenta performs a statistical
analysis of metrics collected from a canary cluster and compares it to a baseline cluster, 
in order to determine whether there has been a statistically significant impact on any metrics of interest.

ChAP leverages Kayenta to perform an automated analysis of the relevant metrics for a chaos engineering
experiment.  While Kayenta was originally designed for evaluating the health of canary deployments, we found
it to be an excellent match for analyzing the results of ChAP experiments.

\begin{figure*}[tbp]
\includegraphics[width=\textwidth]{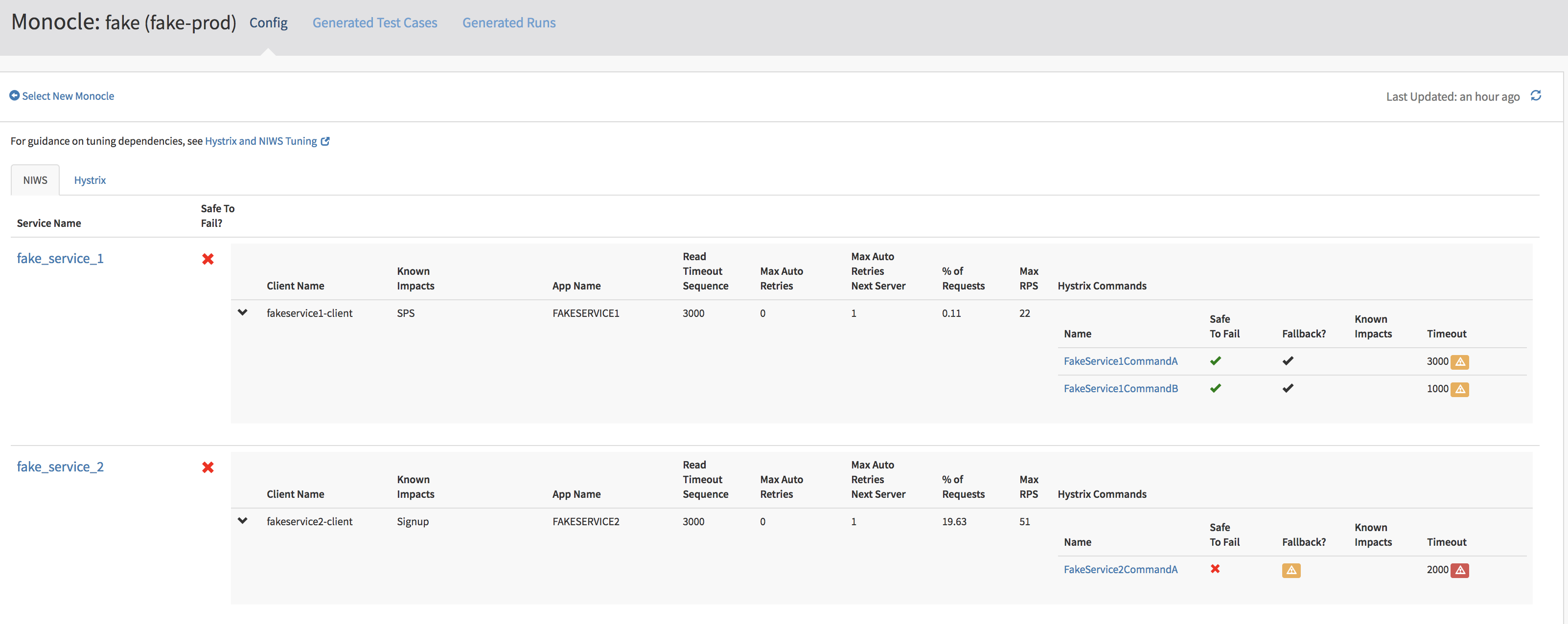}
\caption{Monocle UI for RPC dependencies within a cluster}
\label{fig-niws}
\end{figure*}

\section{Safety mechanisms}
\label{sec-safety}

Because ChAP experiments involve fault injection, every experiment carries risk that it could
lead to an incident. To mitigate this risk, we have built in a number of safety mechanisms
that limit the blast radius of an experiment.

\textbf{Business hours.} ChAP experiments only run during business hours: weekdays
from 9AM to 5PM. This ensures that if something does go
wrong, then engineers are likely to be at work and
can respond more quickly than if there was an issue
after hours.

\textbf{Automatic stop.}  If ChAP detects excessive customer impact during an
experiment, then it will stop the experiment early.
This customer impact test is a cruder test than the
non-parametric statistical tests used by Kayenta,
but it limits the harm in the case where an experiment might have
significant impact on affected users.

\textbf{Total traffic.}  The set of all concurrently executing ChAP experiments cannot impact
more than 5\% of the total traffic, in any one of the three geographical
regions that Netflix servers run in.

\textbf{Failover.} The Netflix control plane is deployed in three different Amazon
geographical regions. If there is a problem with one of the regions, Netflix engineers
can evacuate traffic from the troubled region and redirect it to the other two regions,
a process we refer to as failover\cite{activeactive}. ChAP does not permit experiments
to run during a failover as a safety mechanism, since a failover can violate assumptions
that ChAP makes about the number of requests that are flowing through a region.

\section{Monocle}

In designing ChAP, we considered two models of usage:

\begin{itemize}
    \item a \textit{self-serve} model where engineers define and run their own experiments.
    \item a \textit{fully automated} model where a centralized team defines and runs experiments.
\end{itemize}

We initially adopted a self-serve model, where users were responsible for defining their
own experiments. ChAP integrated with the Spinnaker deployment system so that users could
add ChAP experiments to their deployment pipelines, but users were responsible for setting
this up themselves. Later, we transitioned to a hybrid approach where we would
automatically generate and run experiments, as well as support users in running self-serve experiments.

We developed an additional service called \textit{Monocle} that has two functions, introspecting
services and generating experiments.

\section{Service introspection}

Monocle introspects Netflix services to collect information about its dependencies.
Here, a \textit{dependency}
refers to either a configured RPC client or a Hystrix command.

Monocle integrates data from multiple sources: the telemetry system (Atlas\cite{atlas}), the tracing system (based on Dapper\cite{dapper}) and by querying running servers directly for configuration information such as timeout values.  

Monocle provides a UI which summarizes information about dependencies. For each Hystrix command, Monocle displays:

\begin{itemize}
    \item The percentage of inbound requests that trigger an invocation of the Hystrix command.
    \item Whether it is believed to be safe to fail
    \item Whether it is configured with a fallback
    \item Configured timeouts
    \item Observed latencies over the past two weeks (mean, P90, P99, P99.5)\footnote{P90 stands for ``90\textsuperscript{th} percentile''}
    \item Thread pool size
    \item Observed number of active threads over the past two weeks
    \item Which RPC clients it wraps, if any
    \item Any known impacts associated with the Hystrix command
\end{itemize}

The Monocle UI provides a tabular view from Monocle of a list of Hystrix commands 

We use heuristics such as the presence of a configured fallback and telemetry data
that shows that the fallback has succeeded when executed. When accessing this view, users have the ability to toggle back-and-forth between viewing all of the Hystrix commands for a cluster or viewing all the RPC dependencies for a cluster.

Yellow and red attention icons show tooltips when moused over, and provide additional
details around warnings and vulnerabilities. For example, mousing over red tooltip reveals the following message:

\begin{quote}
Warning: Hystrix Command Timeout is Misaligned with RPC client. 

Timeout (1000 ms) is less than the max computed timeout of the wrapped RPC client (4000 ms).
This means that Hystrix will give up waiting on RPC.  This may be OK for non-critical calls,
but you should review the config settings to confirm the desired behavior.
\end{quote}

%

For each RPC Client, Monocle shows:

\begin{itemize}
    \item Timeout and retry configurations
    \item The percentage of inbound requests that trigger an RPC call
    \item Maximum observed invocation rate (requests per second) over the past two weeks 
    \item Which Hystrix commands wrap it, if any, and if they are safe to fail
    \item Any known impacts associated with the RPC client or the Hystrix commands that wrap it
\end{itemize}

Fig.~\ref{fig-niws} shows a tabular view from Monocle of a list of RPC dependencies (note: NIWS Client reference in the figure is Netflix Internal Web Service Framework, referred to as RPC Client throughout the paper [20]). Note how the commands that are not configured with fallbacks are shown as not being safe to fail. We use heuristics such as the presence of a configured fallback and telemetry data that shows that the fallback has succeeded when executed. When accessing this view, users have the ability to toggle back-and-forth between viewing all of the Hystrix commands for a cluster or viewing all the RPC dependencies for a cluster.

\section{Experiment generation}

Monocle is also responsible for generating experiments automatically. Today, 
Monocle generates fault injection experiments for dependencies (RPC clients and Hystrix
commands). It generates three types:

\begin{itemize}
    \item Failure
    \item Latency just below configured timeout (highest timeout - P99 latency over the past 7 days  + 5\% buffer)
    \item Latency causing failure (highest configured timeout + 5\% buffer)
\end{itemize}

Monocle uses heuristics to try to identify the experiments with the highest
likelihood of finding a vulnerability.

\subsection{Criticality score}

Monocle assigns a \textit{criticality score} to each dependency, which is used as input
for prioritizing experiments. We are interested in prioritizing experiments on the more critical dependencies sooner and more frequently because these are the ones that we believe will cause the most harm if they behave incorrectly. 

Beyond direct experimentation; we have also discussed using scores to generate a list of critical dependencies. That list could be used to identify and prioritize reliability work for our most critical components.

The criticality score is the product of the following four values:

\begin{enumerate}
    \item dependency priority (RPC client $\rightarrow 1$, Hystrix command $\rightarrow 100$)
    \item maximum percentage of inbound requests that trigger the dependency over the past 7 days as compared to all of the dependencies in the cluster ($<0.1\% \rightarrow
          0, <1\% \rightarrow 10, <10\% \rightarrow 100, \leq100\% \rightarrow 1000$)
    \item retry factor (1 + configured number of retries)
    \item number of interactions associated with the dependency (if the dependency is an RPC Client, this would be the number of Hystrix commands, if it is a Hystrix command, this would be the number of RPC Clients)
\end{enumerate}

Hystrix commands are prioritized over RPC clients because Hystrix commands wrap RPC client calls, and we want to verify that the Hystrix fallbacks are working correctly first.

\subsection{Prioritization score}

Monocle calculates a score for each experiment as a the product of:

\begin{enumerate}
    \item criticality score (see above)
    \item safety score (safe $\rightarrow +1$, unsafe $\rightarrow -1$)
    \item experiment weight\footnote{This is a simplification, we actually flip the weights if the score
    is negative in order to maintain the ordering of the experiments.}
    (failure $\rightarrow 3$, latency $\rightarrow 2$, latency causing failure $\rightarrow 1$)
\end{enumerate}

Monocle assigns a safety score of -1 to any experiment that it deems as unsafe. This includes situations such as:

\begin{itemize}
    \item Dependency being experimented on has been blacklisted (no experiment types should be created for this dependency)
    \item Dependency data associated with the experiment has been not been recently collected, and does not have up-to-date information
    \item Dependency being experimented on contains a critical vulnerability and no experiment types should be created. For example, the dependency is an unwrapped RPC Client (no Hystrix command)
    \item Dependency has a known impact associated with its failure (added by a user and directly related to a key performance indicator: SPS, downloads per second, login, or signup)
    \item Experiment being created is a latency experiment, and the dependency is both missing fallbacks and requires timeout tuning 
    \item Experiment being created is a failure experiment and the dependency is not safe to fail due to a missing fallback
\end{itemize}

Monocle computes scores for all possible experiments and the resulting score represents whether a test is safe to run (i.e., scores $>$ 0 are safe). For the current use cases, there is not much need for scores of unsafe experiments except for debugging. This is an implementation detail, and another option would be to pre-process the experiments and identify safety before scoring them.

Monocle then runs the experiments in priority order. Only experiments which have a positive score will be run. When going through the prioritized list, Monocle checks if the experiment should run by ensuring
\begin{itemize}
    \item The experiment is not already running
    \item The experiment is not in a failed state (it has previously failed and has not been looked at by a user)
    \item The experiment has not been run in the last (configured number) of days
\end{itemize}

\section{Results}

\begin{figure}[tbp]
\centerline{\includegraphics[width=\columnwidth]{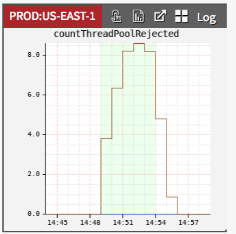}}
\caption{A Hystrix metric graph that shows elevated threadpool rejections for the canary group}
\label{fig-hystrix}
\end{figure}

To date, experiments generated by Monocle have revealed several cases where timeouts were
set incorrectly and fallbacks revealed the service to be more business critical than the owner had intended.  Fig.~\ref{fig-hystrix} shows an example of an interaction problem that can arise between a Hystrix command's
timeout configuration and its threadpool size. In this experiment, about 900 ms of latency is injected into a Hystrix 
command to mimic the scenario of a downstream dependency going latent.

The increase in the \textit{countThreadPoolRejected} metric indicates that
the work could not be scheduled to a thread because all of the threads in the threadpool are blocked, resulting in the Hystrix command serving a fallback. This leads to a short-circuiting behavior where the Hystrix command unconditionally serves fallbacks for a period of time.

The revealed problem is that the timeout is too high relative to the size of the threadpool. In this particular
case, the configured timeout was much higher than the 99\textsuperscript{th} percentile latency reported by our telemetry system, and so the solution was to decrease the timeout accordingly.

\section{Challenges and lessons learned}

\subsection{Failure modeling}

The existing tooling limits the type of faults that we can inject into the system.
In particular, FIT can only cause one type of error per injection point,
but there are often multiple ways that a service can fail. For example, we had a situation where we ran
an experiment by injecting failure into a particular service. The injected fault behaved as if the service 
returned an error. The experiment showed that the system could handle this failure gracefully. However, we had 
an incident where none of the servers associated with that service were registered in our service discovery mechanism.

\subsection{Application-based fault injection}

Netflix implements fault injection using defined injection points in Java-based platform libraries.
These libraries are common to all Netflix applications. Deploying new libraries can take many months, as we have to wait for all of the services to pick them up.

Historically, applications in the Netflix control plane have been Java-based. However, the trend inside of Netflix is to move towards more support for polyglot, most notably JavaScript running on Node.js. This makes an application-based
approach challenging because we need to implement new libraries for each supported language.

There are alternative approaches which do fault injection out-of-process, obviating the need
for language-specific bindings. Istio\cite{istio} is one example of a service mesh based approach
to fault injection. 


%
%

\subsection{Even if you build it, they might not come}

When ChAP was first made available for internal users, we only had a few teams making regular use of
the service. While we strove to make the interface as simple as possible for new users, running an
experiment on production traffic is itself a complex task, and there is a limit as to how much
we can possibly simplify the interface to support this. We employed a consulting model with internal teams who would make good candidates for potential users, but we
did not see widespread usage.

\subsection{Challenge of automation}

Our alternate approach to the self-serve model, automatically generating the experiments ourselves, presented its own
set of challenges. We needed to develop heuristics to determine which types of faults could
be injected which were not known to cause customer impact. This was necessarily a smaller space of 
faults then if we have domain knowledge about the individual services.

In addition, we had to ensure that we had a low false positive rate, otherwise the service
teams would lose confidence in the results of the experiments. This meant that if an automatically
generated ChAP experiment revealed a potential vulnerability, a member of the Resilience Engineering team 
had to spend time analyzing the results of the experiment to verify that it did, in fact, reveal a
genuine vulnerability. This was generally a time-consuming and tedious process. It is possible to
build supplementary tooling to reduce the effort involved in doing the analysis, but that itself
requires an additional investment.

\subsection{Small sub-populations}

Small sub-populations are a challenge to deal with. Netflix software runs a variety of different
types of devices, and these devices behave differently. A problem might only manifest on, say,
a particular brand of Smart TV. If only a tiny fraction of our userbase watches Netflix
on this type of television, then even if these users can't stream video at all, it's unlikely
to be detected by looking at total SPS success counts. We could potentially oversample from
device types that are less common, but this means that we are increasing the blast radius for
that device type, which we have been reluctant to do.

\subsection{Error counts}

Although intuitively error rates seem like a reliable signal for identifying problems with
the system, we found them to have some undesirable properties for use as an experimental measure. 

Error counts can help with the small sub-population issue, because error counts are generally
quite low, so even a smaller sub-population may end up contributing significantly to the error
rate. However, this creates the opposite problem: a single device in the baseline or canary group
can cause a significant increase in the error count if the device keeps erroring over and over.
This means that if an error-prone device happens by chance to be assigned to a baseline or error
group, it can generate a spurious error signal. 

In particular, for our automatic stop safety mechanism, we had to substantially increase the
threshold which we trigger a stop on the error metrics because of how noisy they were.

\subsection{Value of visualization}

To be able to automatically generate experiments, we needed to build tooling to obtain visibility
into the configuration and observed behavior of RPC clients and Hystrix commands. While this information
has always been available within Netflix through telemetry or configuration endpoints, it had not
previously been aggregated and displayed, which proved useful for surfacing vulnerabilities and interactions even without using this
data for automatic experiment generation. For example, by integrating this information into a single view,
we were able to identify some cases of inconsistently configured Hystrix and RPC client timeouts without
even needing to run a ChAP experiment.

\section{Conclusion}

Our work demonstrates that it is possible to automatically and safely generate and run chaos experiments.
These experiments have identified vulnerabilities that could lead to outages if left untreated.

While the original goal for ChAP was to run fault injection experiments, we have discovered
that the platform itself can be used for other types of experiments. 
We are currently extending ChAP to support load testing experiments, similar in spirit to RedLiner\cite{redliner}.
In addition, some self-serve users have begun experimenting with using ChAP for canary
deployments\cite{release-it}, because of the additional analysis that ChAP provides.

\bibliographystyle{IEEEtran}
\bibliography{refs}

\end{document}